\documentstyle[mnras_cite]{mn}
\title[The orientation of the Seyfert nucleus in Markarian 348]
{The orientation of the Seyfert nucleus in Markarian 348\\}
\author[S.~Ant\'on, A.~Thean, A.~Pedlar, IWA.~Browne]
{S.~Ant\'on$^{1,3}$\thanks{Guest User,
Canadian Astronomy Data Center, which is operated by the Dominion
Astrophysical Observatory for the National Research Council of
Canada's Herzberg Institute of Astrophysics.}, A.H.C. Thean$^{1,2}$, A. Pedlar$^1$, I.W.A. Browne$^1$\\
$^1$~Jodrell~Bank~Observatory, University of Manchester, U.K.\\
$^2$~Istituto di Radioastronomia-CNR, Bologna, Italy\\
$^3$~CAAUL,Observat\'orio Astron\'omico de Lisboa, Portugal\\}

\begin{document}
\input{psfig}

\label{firstpage}
\maketitle

\begin{abstract}
Multi-frequency observations of Mrk 348 are presented consisting of 2 epochs of
MERLIN data at 5 GHz, ISOPHOT data at 170, 90, 60 and 25 $\mu$m, NOT
images at U, B, R and I bands and data at 1350$\mu$m from the archive
for SCUBA.  The new optical images reveal a disrupted arm that ends
towards the eastern companion of Mrk 348, consistent with the
hypothesis that Mrk 348 and its companion form an interacting system.
5 GHz MERLIN imaging shows that only one of the radio components of
Mrk 348 is polarized (\%P=5).  The broadband spectrum of Mrk 348 is
flat between the radio and millimetre bands, suggesting that
synchrotron emission extends to high frequencies.  Mrk 348 has many of
the characteristics of a radio-loud object. 
We discuss the orientation of the radio axis of Mrk 348
with respect to the line of sight. We conclude that the evidence is
conflicting, and the geometry in Mrk 348 is not well-described by a
simple edge-on or face-on model.

\end{abstract}


\section{Introduction}

\noindent  Active Galactic Nuclei (AGN) are thought
of as existing in broadly two flavours: those with strong radio emission, 
the ``radio-loud'', and those with weaker radio emission, the 
``radio-quiet''. There is strong 
evidence that radio-loud AGNs are hosted by  elliptical galaxies, 
whereas radio-quiet AGNs, at least those of low redshift,  are mainly found
in spiral galaxies. Recently, however, the difference between radio-loud 
objects and radio-quiet has been blurred by results that show that at
high redshift (and  high luminosity) the hosts of radio-quiet quasars 
are ellipticals too \footnotemark\footnotetext{Apart
from few exceptions (e.g. III Zw 2), powerful radio-quiet AGNs have not been 
detected in spiral galaxies.} (e.g. \pcite{mclure99}).
Apart from radio emission
(and related non-thermal emission at other wavelengths), there
is almost no difference  between radio-quiet and radio-loud sources of the
same optical spectroscopic type. \\
\noindent Mrk 348 (B0046+316) is a relatively nearby (z=0.015) bright
Seyfert 2 galaxy (R=13 mag) that has been extensively studied.  The
optical total intensity spectrum of Mrk348 shows narrow emission lines
(e.g. March\~a et al. 1996) while spectropolarimetric observations
reveal a hidden broad line region (\pcite{tran95}, \pcite{MG90}).
Observed by the VLA\footnotemark\footnotetext{Very Large Array} in A
configuration at 8.4 GHz, Mrk348 has a core-dominated unresolved radio
morphology \cite{wiletal98}.  At 5 GHz the radio nucleus of Mrk348 at
VLBI\footnotemark\footnotetext{Very Long Baseline Interferometry}
resolution has a linear triple structure with 0.2$''$ size at
a PA=168$^o$ \cite{neff83}. Capetti et al. (1996) detected [OIII] emission
confined in a linear structure with 0.45$''$ size that aligns very
well with the triple radio structure. Also, Falcke et al. (1998) found that
the emission-line region is very concentrated within the central arcsecond.
 MERLIN\footnotemark\footnotetext{Multi Element Radio Linked
Interferometer Network} observations show that the central component
of the triple has an inverted radio spectrum, suggesting it is the
core \cite{unger84}. Flux density monitoring shows that the core is
variable at 5 GHz on scales of months \cite{neff83}. At 15 GHz and
with the VLBA\footnotemark\footnotetext{Very Long Baseline Array}
resolution of $\sim$1~mas, Mrk 348 shows a one sided core-jet structure
(Ulvestad et al. 1999; hereafter U99). The comparison of two epochs of VLBA
observations suggests that the radio components are expanding with
sub-relativistic speeds (U99). Between 1997.1 and
1998.75 the core has undergone a major radio flare, increasing by a
factor of 5.5 in flux density (U99). {\it Ginga}
observations of Mrk 348 reveal a hard X-ray component, which is
absorbed at low energies; the estimated line-of-sight column density
is n$_{\mbox \footnotesize H}\sim 10^{23.1}$cm$^{-2}$ (e.g. Warwick et
al. 1989). Recently Falcke et al. (2000) reported the detection of an
H$_2$O megamaser in Mrk 348.\\
\noindent Mrk 348 has been usually discussed in studies concerning the 
properties of Seyfert galaxies, which are mainly radio-quiet objects.
However, the radio properties of  Mrk 348 are comparable with those of 
low-luminosity radio-loud objects.
Mrk 348 challenges the standard
radio-loud/radio-quiet picture in that it is a relatively strong
radio emitter but it is hosted by a spiral galaxy.
Some of the properties of Mrk 348 are very similar to those of III Zw 2, a 
Seyfert I galaxy in which superluminal motion has recently been detected
\cite{Brunthaler00}.
Even though no superluminal motion has been observed in Mrk 348, its
properties fit nicely the expectation for a radio-intermediate 
quasar (as it is the case of III Zw 2), of the kind suggested by Falcke et al.
(1996).
 In this scenario
the objects are intrinsically weak radio emitters but, 
due to relativistic boosting, their radio emission is amplified.\\
We have been gathering new data on Mrk 348, which are presented and 
discussed in this paper. The new set of observations comprise multi-frequency
 data from the radio to the optical band. There are 2 epochs of MERLIN data at 
5 GHz, ISOPHOT\footnotemark\footnotetext{ISOPHOT, one of the 4 instruments 
onboard of the Infrared Space Observatory (ISO)} data at 170, 90, 60 and 25 $\mu$m, and high (ground-based 
telescope) resolution NOT\footnotemark\footnotetext{Nordic Optical Telescope}
 images at U, B, R and I bands. We also present
data at 1350$\mu$m from the JCMT\footnotemark\footnotetext{the James Clerk Maxwell Telescope} archive for 
SCUBA\footnotemark\footnotetext{Submillimetre Common-User Bolometer Array
on the JCMT.}. 
In this paper
we discuss the host galaxy of Mrk 348, the spectral energy distribution (SED)
of its nuclear emission and the polarization at 5 GHz. Based
on the available information,  we discuss  the orientation of the radio axis 
of Mrk 348 with respect to the line of sight. The paper is organised as 
follows: in Section~\ref{section-obs} the 
multi-frequency observations and data reduction are presented, in 
Section~\ref{largescale} both the large-scale properties and the
central region properties of Mrk 348 are 
discussed. In Section~\ref{mrk348radiodiscussion} we discuss the 
radio-loudness of Mrk 348, including the orientation of 
the radio emission with respect to the line of sight. The important points 
of this work are summarised in Section ~\ref{section-conclusions}.
Throughout the paper we assume H$_o$=65kms$^{-1}$Mpc$^{-1}$ and
$q_o$=0.

\section{Observations and data reduction}
\label{section-obs}

\begin{figure*}
\centerline{
\psfig{figure=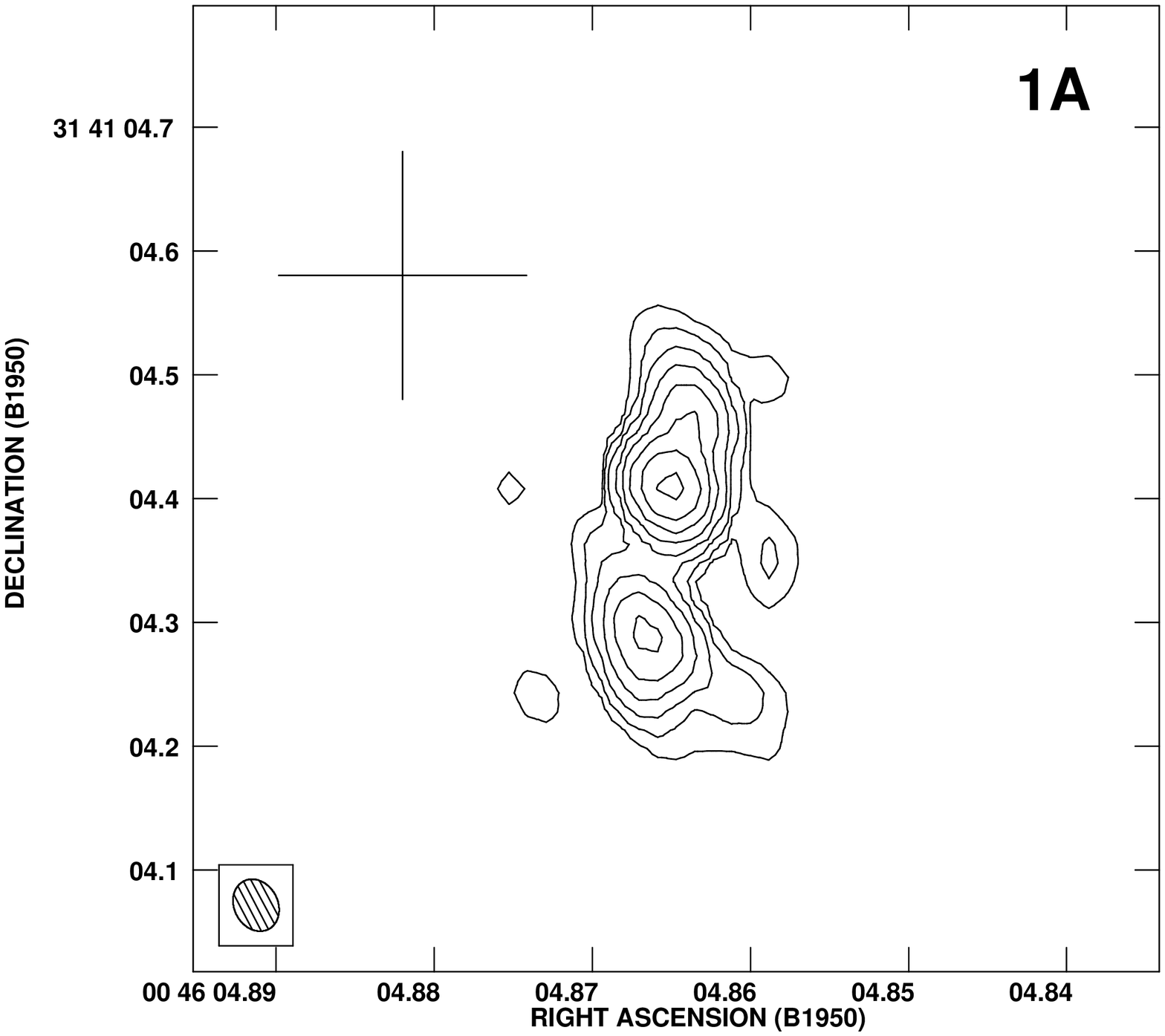,width=7cm,angle=-90}
\psfig{figure=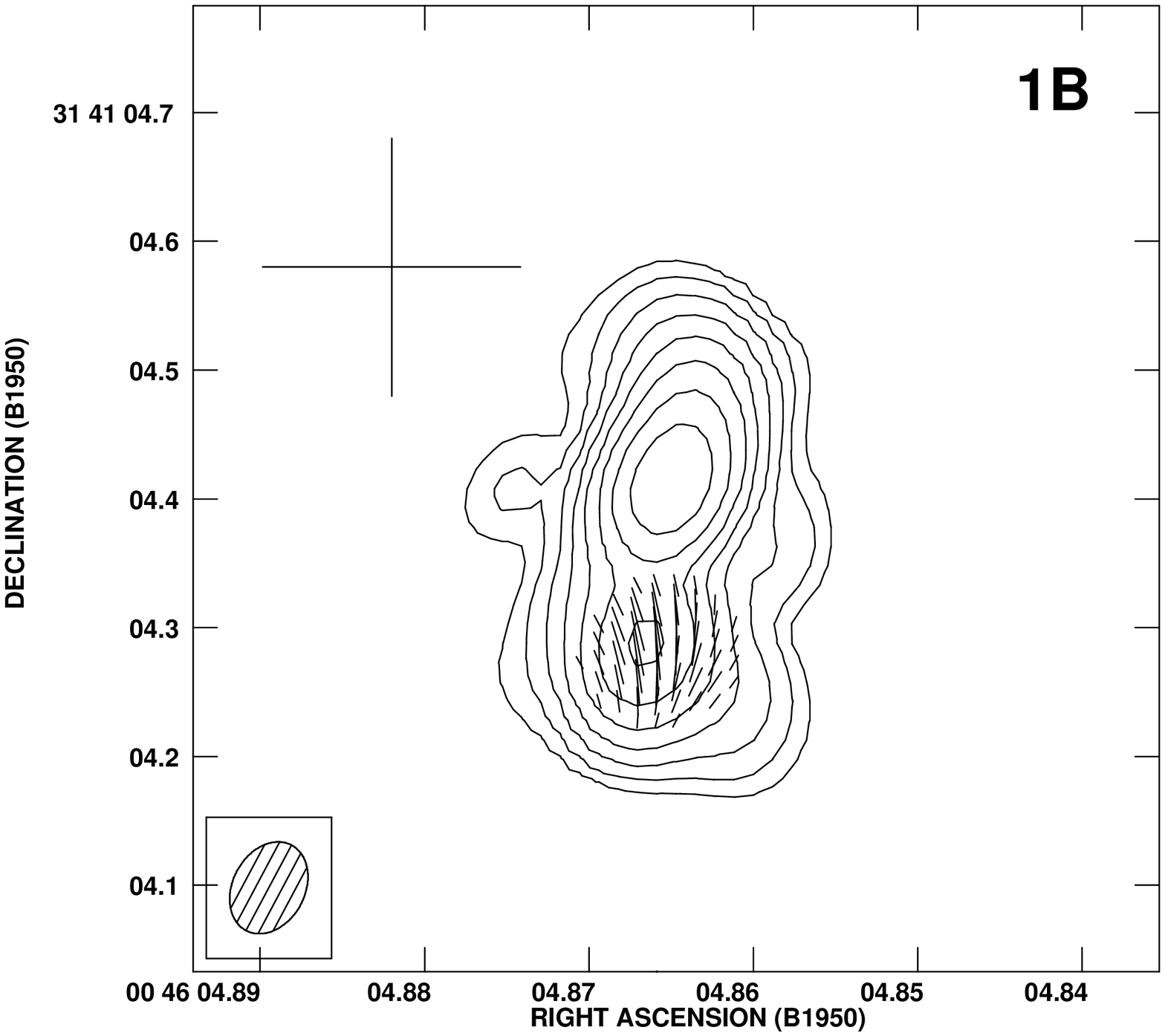,width=7cm,angle=-90}}
\centerline{
\psfig{figure=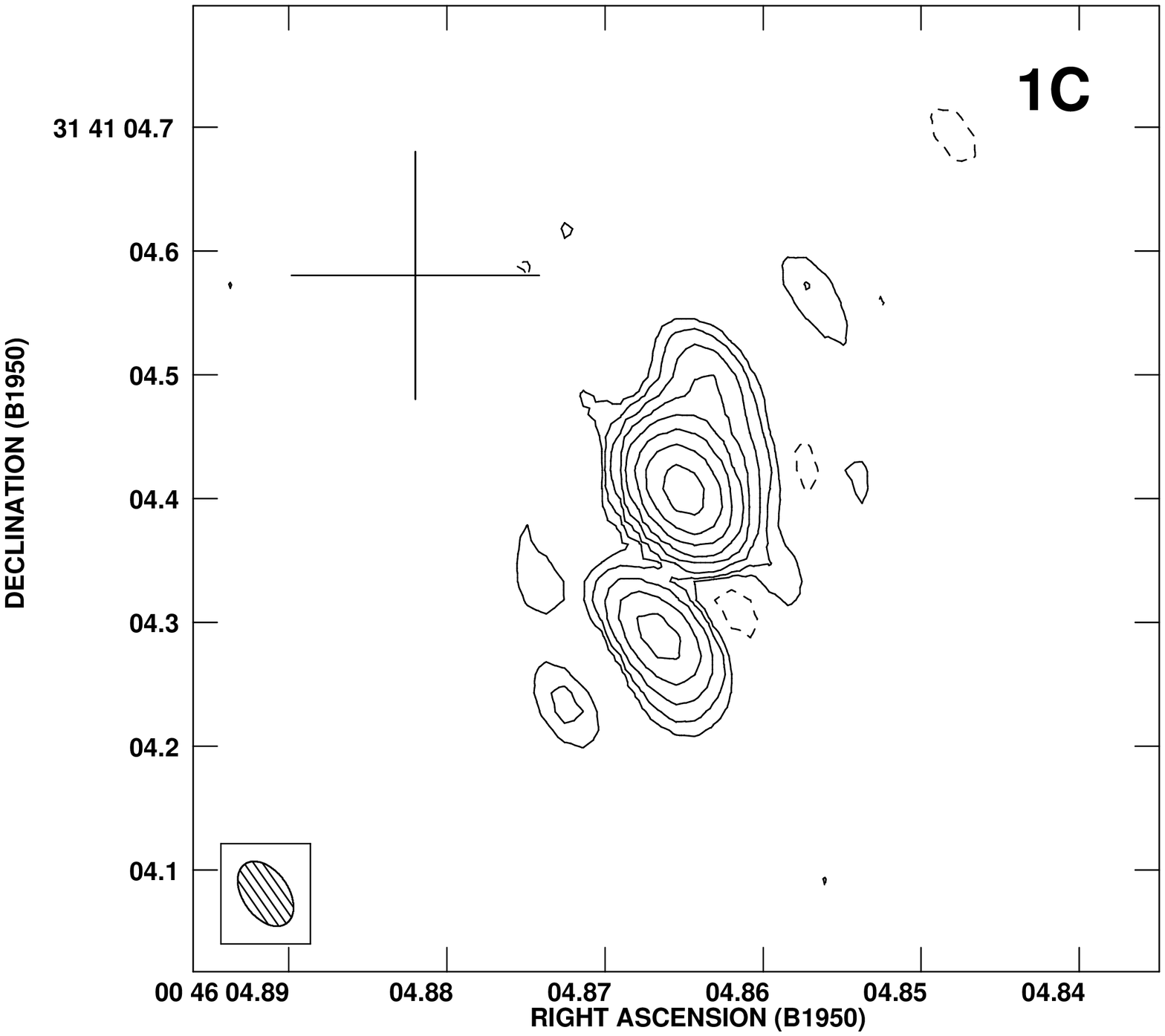,width=7cm,angle=-90}
\psfig{figure=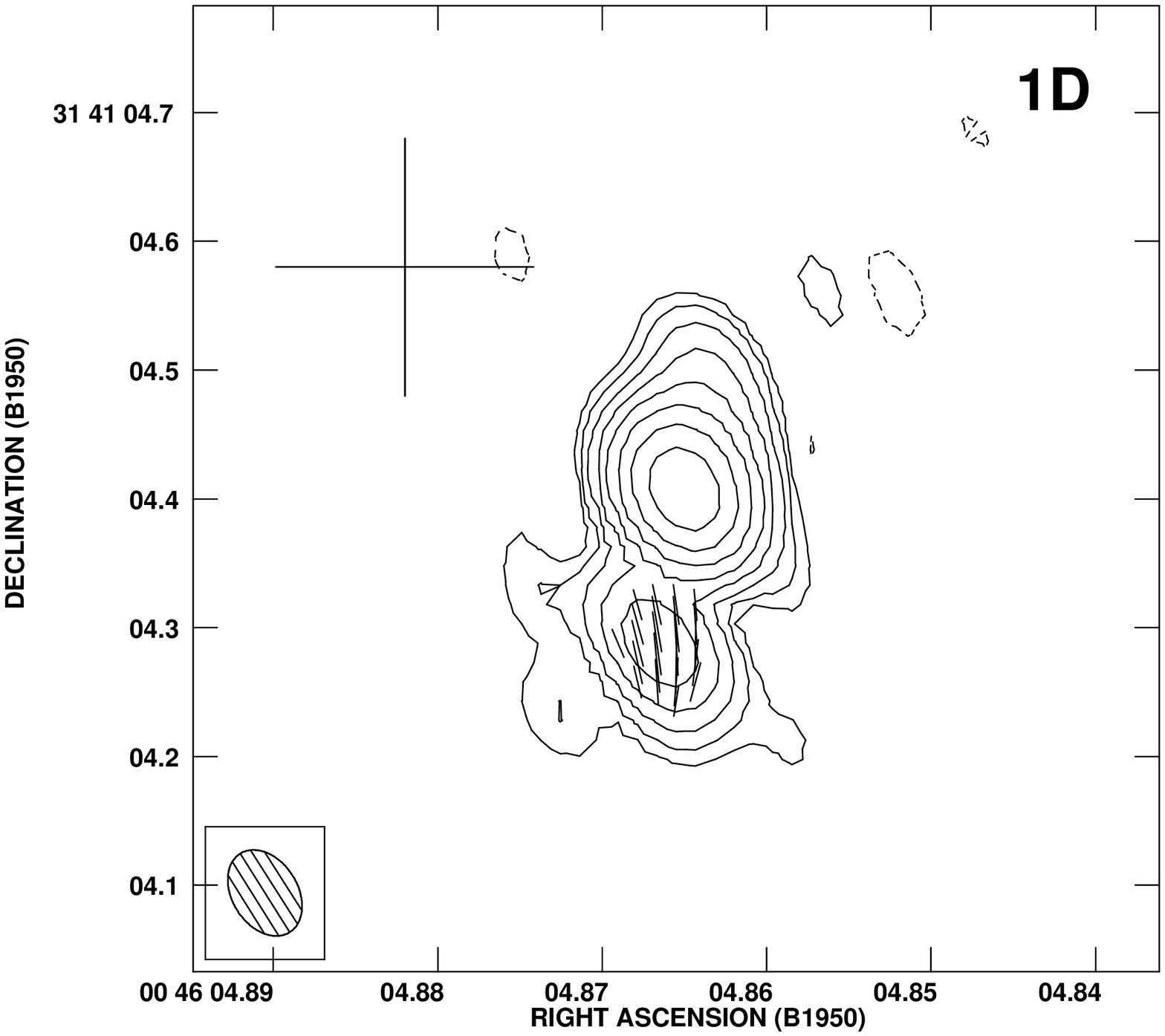,width=7cm,angle=-90}}
\caption{MERLIN maps at 5 GHz showing the 1992 uniformly--weighted map (1A),
the 1992 naturally--weighted map (1B), the 1998 uniformly--weighted map (1C)
and the 1998 naturally--weighted map (1D).
The root--mean--square noise levels for each map ($\sigma$), as measured in
an un--CLEANed region at the edge of the field, are; 0.175 mJy/beam (1A), 
0.135 mJy/beam (1B), 0.177 mJy/beam (1C) and 0.158 mJy/beam (1D).
Contour levels are chosen as 3$\sigma\times$(-2,-1,1,2,4,8,16,32,64,128)
for the 1992 maps and 6$\sigma\times$(-2,-1,1,2,4,8,16,32,64,128) for the
1998 maps.
Crosses show the optical position of the nucleus as given by 
Clements (1981); the size of the cross shows the quoted positional uncertainty.
Polarisation vectors with a scale of 1 mJy/beam to 0.015 arcsec 
are overlayed on the naturally--weighted maps.}
\label{maps.fig}
\end{figure*}

\subsection{MERLIN data}
\label{section-merlin}
We obtained MERLIN 5--GHz images of Markarian 348 on two occasions. 
The first observing run took place on the 14th of December 1992,
the second on the 10th of November 1998. A bandwidth of 15 MHz, centred at
4993 MHz, with both right and left polarisation, was used. 
Flux density calibration was carried out using 3C 286 and relative baseline
gains were determined by observing OQ 208. 
For each run, the data were phase calibrated using different nearby
calibrators; 0103+337 (1992 observations) and 0052+298 (1998 observations). 
In order to register images from the two epochs we have 
introduced a +0.1398 arcsec shift in RA to the 1992 data (the
difference in position between the positions of the southern component 
in the uniform maps); 
the position assumed for 0052+298 ($\alpha_{J2000}$=00h 54m 45.88973s,
$\delta_{J2000}$ = 30$^\circ$ 06$'$ 58.3828$''$), as taken from 
Wilkinson et al. (1998), 
is probably more reliable than that assumed for
0103+337 ($\alpha_{J2000}$ = 01h 06m 00.28710s, $\delta_{J2000}$ = 34$^\circ$ 02$'$
03.0392$''$). 
The data were subjected to several passes of self--calibration. 
The resulting data were Fourier--transformed to a
0.015 arcsec grid resulting in angular resolutions of:
a) 1992 epoch -- 73x54 mas using a natural-weighting scheme and 44x35 mas
using a uniform-weighting scheme b) 1998 epoch -- 72x51 mas using a natural
weighting-scheme and 58x37 mas using a uniform-weighting scheme.\\
Naturally--weighted and uniformly--weighted 
contour maps of the source at the two epochs are presented in Figure~\ref{maps.fig}.
Flux densities were determined using the AIPS task TVSTAT and the
positions measured using JMFIT. The flux densities and positions derived
from the uniform images are given in Table \ref{tabMERLIN}. The 1992 and 1998
measurements are consistent with each other within the errors, although
the flux density of the central component has varied by approximately a
factor of two. The relative positions of the components at the two epochs agree
within 5 mas. The images are consistent with the lower resolution (70 mas)
early MERLIN results at 5 GHz (Unger et al. 1984) and the VLBI measurements
of Neff \& deBruyn (1983).\\
The images were calibrated for polarisation using 3C 286; the
polarisation vectors are shown in Figure~\ref{maps.fig}. Only the
southern component shows significant polarisation and the direction of
the polarisation vectors are approximately North-South. The
polarized flux density of the southern component was 2.5 mJy and 3 mJy in
the 1992 and 1998 data sets respectively. When noise and sources of
calibration errors are taken into  account the increase in polarised flux  
is not significant. Hence the southern component is  $\sim$5
\% polarised, whereas the stronger central component shows no polarisation
greater than 1 mJy ($<$0.5 \%).

\begin{table}
{\footnotesize
\begin{tabular}{ccccr}
 \hline \hline
\multicolumn{1}{c}{Component} &
\multicolumn{1}{c}{Epoch} &
\multicolumn{1}{c}{RA} &
\multicolumn{1}{c}{Dec}   & 
\multicolumn{1}{c}{F$_{\mbox{5GHz}}$}    \\ 
\multicolumn{1}{c}{} &
\multicolumn{1}{c}{} &
\multicolumn{1}{c}{{\scriptsize (B1950)}} &
\multicolumn{1}{c}{{\scriptsize (B1950)}}   & 
\multicolumn{1}{c}{{\scriptsize [mJy]}}  \\
\hline 
& & & & \\
Radio Core   &  1992 & 00 46 04.86501 & 31 41 04.4118 & 116.4\\
Southern     &   ''  & 00 46 04.86653 & 31 41 04.2880 &  40.5\\
& & & &\\
Radio Core   & 1998  & 00 46 04.86499 & 31 41 04.4075 & 248.1\\
 Southern   &   ''  & 00 46 04.86653 & 31 41 04.2879 &  31.6\\
& & & &\\
\hline
& & & &\\
 Total & 1992 &&&  185.0 \\
 Total & 1998 &&&  299.6 \\
& & & &\\
\hline \hline
\end{tabular}
}
\caption{The positions and flux densities of the nuclear radio components
of Mrk 348 as measured with MERLIN at 5 GHz.}
\label{tabMERLIN}
\end{table}

\subsection{SCUBA - Data from Archive}
\label{section-scuba}
Mrk 348 was observed at the JCMT with SCUBA at 1350 $\mu$m and we
retrieved the data from the archive.  The data were processed with the
SURF (SCUBA User Reduction Facility) package. The reduction process
was based in the techniques described in Stevens et al. (1997), which
includes the nod compensation, flatfielding, correction for extinction
and flux density calibration. \\
\noindent We obtained a value for the flux density at 1350 
$\mu$m of 187 $\pm$ 8 mJy, the error on the flux density
includes the calibration uncertainties and the instrumental flux 
uncertainties, added in quadrature. \\

\begin{figure*}
\centerline{
\psfig{figure=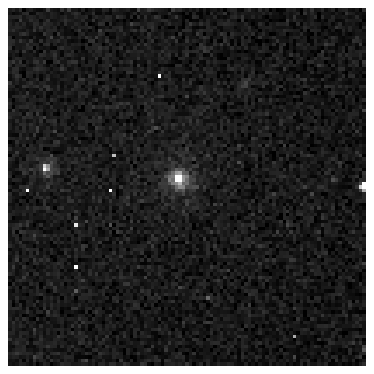,width=7cm}
\psfig{figure=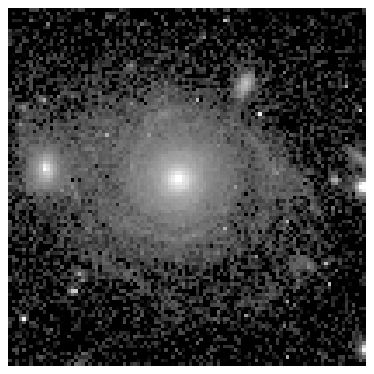,width=7cm}}
\centerline{
\psfig{figure=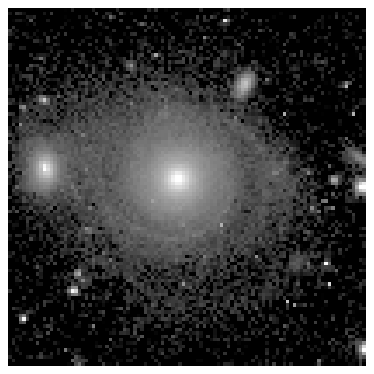,width=7cm}
\psfig{figure=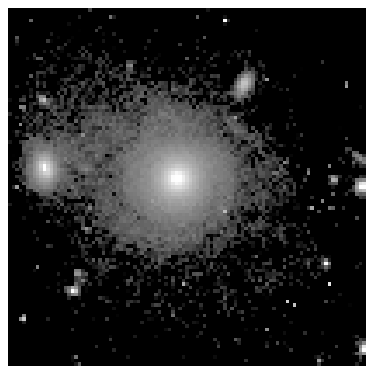,width=7cm}}
\caption{U, B, R and I NOT Mrk348 images. For each image
North is up and East is on the left, the field of view is 3.3$'\times 3.3'$.
{\bf Top Left}: U image of Mrk348. In the image we detect Mrk348 -- central 
object -- and also its eastern companion, a galaxy at the same redshift 
as Mrk348. {\bf Top Right}: B image of Mrk348. This image reveals fine 
structure in the spiral arms: in the inner regions the arms are very 
symmetric, but the external arms show a disrupted morphology.  
The outermost arm ends towards the eastern companion of Mrk348.
{\bf Bottom Left}: R image of Mrk348. This image is similar to the B image,
but we detect the more evolved red stellar population. {\bf Bottom Right}:
I image of Mrk348. This image is similar to R image, but here the external 
arm appears dimmer. Some of the faint features in the I image are 
due to fringing in the CCD.}
\label{notubri}
\end{figure*}

\subsection{ISOPHOT Observations}
\label{section-iso}

\noindent Mrk 348 was observed at 25, 60, 90 and 170 $\mu$m wavelengths by the
Infrared Space Observatory (ISO) with the imaging photo-polarimeter ISOPHOT. 
The observations were performed in PHT03 and PHT22 (P03 and P22 hereafter) 
photometric modes, in chopped mode, with the same integration time on-source 
and on-background. The observations were carried out 
on the 24th of December 1996 and on the 12th of January 1997. The integration
time in P03 mode was of 268 sec, and that in P22 mode was of 374 sec.
P03 observations were taken with triangular chopper mode, 
with a throw of 60$''$ between positions. P22 measurements were performed in 
rectangular chopper mode; the chopper throw was 180$''$ between positions. 
The data were reduced using PIA\footnotemark\footnotetext{PHT-Interactive Analysis. 
PIA is a joint 
development by the ESA Astrophysics Division and the ISOPHOT consortium.}. \\
\noindent Table~\ref{tablelogiso} presents the results from ISO
photometry.  The quoted flux densities are the means of the flux
densities (obtained from the several chopper measurements) and
$\sigma$ is the dispersion of these flux densities about the mean, not
the dispersion on the mean $\sigma$/$\sqrt{\mbox{N}}$. Though for a Gaussian
distribution the best estimate of the error on the mean is
$\sigma$/$\sqrt{\mbox{N}}$, in our case we do not expect a
Gaussian error distribution because the signals showed distortions due
to glitches and/or long term drifts. Thus $\sigma$ gives a
(conservative) estimate of the significance of the measurement. Also,
$\sigma$ does not contain any contribution from systematic errors, as
they were not quantified at the time of the data reduction. In this
context, a detection means that the mean flux density is higher than
3$\sigma$, which is the case for Mrk 348 at the 4 infrared
wavelengths. Mrk 348 had been detected by IRAS, and at 60 $\mu$m there
is good agreement between the IRAS (F$^{60\mu m}_{IRAS}$=$1290\pm 116$
mJy) and ISOPHOT flux densities.  The IRAS 100 $\mu$m flux density
(F$^{100\mu m}_{IRAS}$=$1549\pm 201$ mJy) is slightly higher than the
ISO flux density at 90 $\mu$m.  Much more significant is the disagreement
between the 25 $\mu$m flux densities. The 25 $\mu$m
ISOPHOT flux density  is approximately a factor of 3 smaller than
IRAS flux density (F$^{25\mu m}_{IRAS}$=$835\pm 25$ mJy). We suggest
that this discrepancy might be explained as the sum of two effects: a)
P03 measurements were performed with a 4s/chopper plateau chopper
frequency, and for this setup radiation loss might occur (Haas, priv
comm.) b) the chopper throw of 60$''$ is small compared to the extent
of the disk of the galaxy, and the background position is possibly
contaminated by residual galaxy emission.

\begin{table}
{\center
{\footnotesize
\begin{tabular}{lrrr}
\hline \hline
\multicolumn{1}{c}{F$_{25}$ [mJy]} &\multicolumn{1}{c}{F$_{60}$ [mJy]} & 
\multicolumn{1}{c}{F$_{90}$ [mJy]} & \multicolumn{1}{c}{F$_{170}$ [mJy]} \\
\hline
260 (57) & 1139 (330) & 1108 (302) &  1451 (300) \\
\hline \hline
\end{tabular}
}
\caption{ISO flux densities at 25, 60, 90 and 170~$\mu$; the rms is in parenthesis. The rms is the
dispersion of the distribution of the flux densities (see text).}
\label{tablelogiso}
}
\end{table}

\subsection{NOT data}
\label{section-not}
\begin{figure}
\centerline{
\psfig{figure=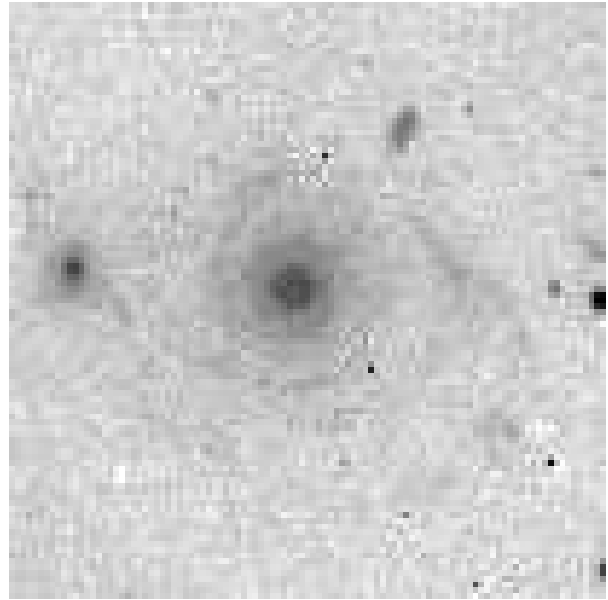,width=8cm}}
\caption{B-I image of Mrk348. North is up and East is
left and the field of view is $3.3'\times 3.3'$. We detect a slight colour
 gradient in the
eastern companion. The external bluish arms of Mrk348 are noticeable, as 
is the spiral structure in the inner regions.}
\label{figbmi}
\end{figure}

In August of 1997 we obtained U, B, R and I images of Mrk 348 with the 2.6~m Nordic Optical Telescope (NOT).
The observations were performed with the HiRAC camera using a 2k Loral CCD, 
with a field view of $3.7\times3.7$ arcmin$^2$ and a pixel scale of
0.11 arcsec. The conditions were photometric with seeing $\sim$ 0.7 arcsec
throughout the observations. 
 Standard stars from \scite{landolt1992} 
were observed and sky flats were obtained at the
beginning and at the end of the nights. The data reduction was performed with 
IRAF. The method followed the standard procedures.
The extinction coefficients were obtained through the standard stars. 
Airmass corrections were applied. The standard calibrations were computed to
obtain the zero point magnitudes. The root mean square error 
of the calibrations is  RMS=0.03 mag for B magnitudes, RMS=0.02 mag for R
magnitudes and RMS= 0.01 mag for I magnitudes.\
\noindent A Point Spread Function (PSF) model was derived independently 
from each image, using {\bf daophot} tasks. The colour image B-I was obtained 
by dividing the I image by the B image. To do this, the I image was shifted 
in position
(to get a perfect matching between the 2 frames); here Galactic stars were 
used to obtain the registration of the frames. Then, the shifted I image was
smoothed with a Gaussian filter in order to match the
resolution of B image; the task {\bf gauss}, which
convolves an image with an elliptical Gauss function, was used for this purpose.\\
In Figure~\ref{notubri} U, B, R  and I images are presented. 
North is at the top and East is on the left. Each field of view is 
3.3x3.3 arcmin$^2$ and 1$''$ represents 327 pc in the rest frame of the source.
 Figure~\ref{figbmi} presents the B-I colour image of 
Mrk348, where darker pixels represent bluer colours and lighter represent 
redder colours.  Table~\ref{colours} lists the B, R and I magnitudes for
different apertures centered on the nucleus. \\

\par

{\footnotesize
\begin{table}
\label{tab}
\begin{center}
\begin{tabular}[t]{ccccc}
 \hline \hline
\multicolumn{1}{c}{Aperture [$''$]} & \multicolumn{1}{c}{}  & \multicolumn{1}{c}{B [mag]}
  & \multicolumn{1}{c}{R [mag]}   & \multicolumn{1}{c}{I [mag]}    \\ 
\hline 
 1.0   &  &  18.07 &  16.60 & 16.04 \\ 
 1.5   &  &  17.59 &  16.12 & 15.53 \\
 2.0   &  &  17.42 &  15.95 & 15.35 \\
 3.0   &  &  16.86 &  15.39 & 14.74 \\
 6.0   &  &  16.15 &  14.68 & 13.97 \\
 9.0   &  &  15.78 &  14.32 & 13.59 \\
12.0   &  &  15.52 &  14.10 & 13.36 \\
\hline \hline
\end{tabular}
\end{center}
\caption{Photometry (NOT images) of Mrk348 through 
different apertures.}
\label{colours}
\end{table}
}
\par

\section{Results}
\label{largescale}
\subsection{Large-scale properties}

\begin{figure}
\centerline{
\psfig{figure=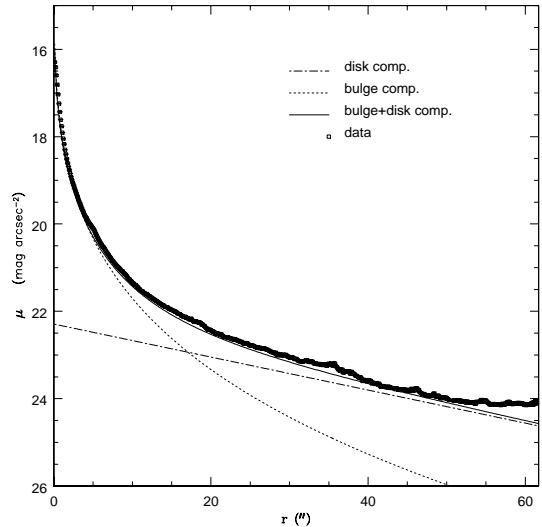,width=8cm}}
\caption{Bulge-disk decomposition of Mrk 348 brightness profile.}
\label{figbulgedisk}
\end{figure}
\par

\noindent Broadband optical images of Mrk 348 have been obtained with
the Hubble Space Telescope and show the structure of the inner regions
of the galaxy \cite{mal98,cap96}. The advantage of NOT imaging is
(with good seeing, as in our case) the ability to reveal structure in
the low surface brightness regions, adjacent to the HST
structure. This is well demonstrated by our B image (see
Figure~\ref{notubri}) that reveals fine details of the spiral
arms. Our images confirm that Mrk348 is a giant face-on spiral
galaxy. From the B image we estimate an inclination angle of $\approx$
20$^o$. In the past the galaxy morphology has been classified as S0
\cite{Huchra80}, but our NOT images suggest that the morphology
corresponds to a later-type.  We have analysed the brightness profile
of Mrk 348, which is presented in fig.\ref{figbulgedisk}.  The details
of the method used to compute the profile are given in March\~a et al.
(in preparation). Briefly, the profile was obtained using the IRAF
package {\bf ellipse}, where elliptical isophotes were fitted to the
galaxy image down to a surface brightness of $\mu= 24.5$ mag
arcsec$^{-2}$. A radial profile of the average isophotal surface
brightness versus semimajor axis was then produced and model light
distributions were fitted to the profile using a non-linear
least-squares method.  The best fit to this profile was obtained using a
function that is the sum of a Vaucouleurs law and a full exponential
disk, I($r$)= I($r$)$_{bulge}$+I($r$)$_{disk}$ (note $\mu \sim -2.5
\log$ I), where:\\

\centerline{I($r$)$_{bulge}$ = $ I_e$ $e^{-7.688}$ $^{[ (\frac{r}{r_e})^{\frac{1}{4}} - 1.0 ]}$}
\centerline{I($r$)$_{disk}$=$I_o$ $e^{-}$ $^{(\frac{r}{r_o})}$.}

\begin{figure}
\centerline{
\psfig{figure=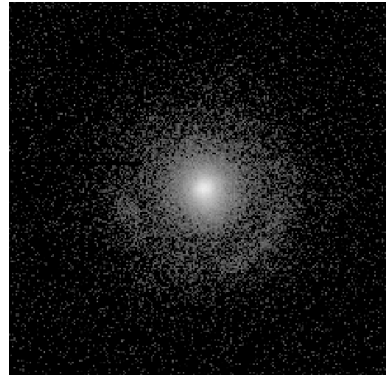,width=5cm}}
\caption{Central U image of Mrk348. North 
is up and East is left and the field of view is 24$'' \times 24''$ i.e. 
$\sim 10 \times$10 kpc$^2$.
Pseudo-colour image with logarithmic brightness scale. 
The ring-like distribution coincides with HII regions.}
\label{fignucU}
\end{figure}

\noindent $r_e$ is defined as the radius within which 
half of the total light is emitted, $I_e$ is the intensity at $r_e$,
$I_o$ is the central intensity of the 
disk, $r_o$ is the disk scale length. The parameters of the best fit are 
$r_e= 2.88 \pm 0.01$ kpc, $\mu_e = 21.44 \pm 0.01 $ mag arcsec$^{-2}$, $r_o= 9.39 \pm 0.02$ kpc,
$\mu_0 = 22.29 \pm 0.01$ mag arcsec$^{-2}$. The bulge-fraction, defined as 
B/T = $\frac{r_e^2 I_e}{r_e^2 I_e+0.28r_o^2 I_o}$, and related to the 
disk-to-bulge ratio D/B by B/T = 1/(D/B+1), is 0.4. According with Fig. 4.51 
in Binney \& Merriefield (1998) this value indicates that Mrk 348 lies between 
 an Sa+ and an Sb+ type spiral galaxy. Our data demonstrate that the morphology of Mrk 348 host is indeed 
of a later-type than S0.\\
Our images show that a disrupted spiral arm ends towards the
eastern companion of Mrk 348, 
suggesting that Mrk348 and its eastern companion form an interacting 
system, as first proposed by \scite{sim87} primarily on the basis of radio 
observations of the distribution of neutral hydrogen. 
Also, the eastern companion shows a colour gradient, being 
bluer near the region where the external arm of Mrk348 seems to end (see 
Figure~\ref{figbmi}). If the bluish colour is the result of
star forming regions, star formation could have been triggered by tidal
interaction between Mrk348 and its eastern companion. Perhaps the interaction
between Mrk348 and its eastern companion is responsible for triggering the 
nuclear activity of Mrk 348, as suggested by Simkin et al. (1987). We note, 
however, that the spiral structure 
on scales up to 30 kpc of Mrk348 is very symmetric, something that contrasts
with the obviously disturbed outer structure. If the interaction is 
responsible for triggering the nuclear activity of Mrk 348 it is surprising 
that the spiral structure has managed to remain so undisturbed.\\
\noindent In Figure~\ref{fignucU} we present an enlargement of the U image 
(Figure~\ref{notubri}) showing the central 
region (24$'' \times 24''$). A  ring of continuum
emission with a diameter of $\sim 3.2$ kpc can be seen. This ring is blue 
(see Figure~\ref{figbmi}) and is coincident
with the H$\alpha$ emission regions detected by \scite{delgado97}, suggesting
that these are HII regions.\\
Motivated by the  Galactic Dust Model (GDM) proposed by \scite{mal98}
we analysed the colour image B-I to try to identify any large-scale dusty 
regions. In the GDM the classification of an object as a Seyfert 1
or a Seyfert 2 depends on how the dusty regions are distributed in the inner
parts  of the host galaxy; if it happens
that a galactic
dusty region  is between the observer and the central engine, then the object
is classified as a Seyfert 2. We do not see any evidence of extra reddening
on scales we can resolve ($>$ 200 pc). We conclude that large-scale 
dusty regions are not obscuring the BLR\footnotemark\footnotetext{
Note that evidence for dust in the vicinity of the very central region has 
been found (see Falcke et al. 1998, Simpson et al. 1996).}.\\
\noindent Comparison of our aperture photometry with that presented in 
Kotilainen \& Ward (1997)  suggests that the central region of Mrk 348 
changed colour between 1992 and 1997 (Ant\'on et. al., in preparation).

\par

\subsection{Properties of the central regions -- The Spectral Energy 
Distribution (SED)}

\begin{figure}
\centerline{
\psfig{figure=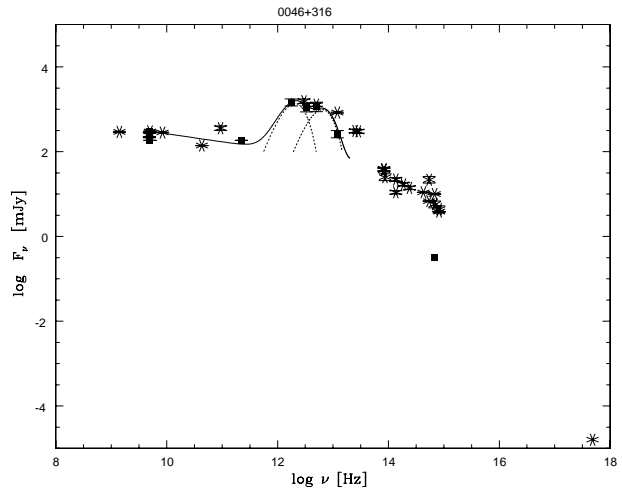,width=7cm,angle=-90}}
\caption{0046+316: Flux densities from the literature are shown in star 
symbols, including IRAS data (12, 25 60 and 100 $\mu$m). The data presented 
in this paper are represented by square symbols. The solid line represents the sum of 
a power law spectrum  with two greybody spectra of temperatures 
$\sim$ 20 K and $\sim$ 62 K (each component is represented by dashed lines).}
\label{iso0046}
\end{figure}

The SED, spanning wavelengths from the radio to the X-ray bands, is presented
in Figure~\ref{iso0046}. Comparing the emission at radio wavelengths to 
that at the ISO wavelengths it is clear that the spectrum shows an excess 
at the infrared wavelengths, and this excess is most naturally explained by 
emission by dust probably heated by the AGN. We fitted the SED between radio 
and infrared bands with 
a power-law component plus two greybody components, a cooler one with 
temperature $\sim$ 20 K and a warmer one with temperature $\sim$ 62 K. 
\noindent  Mrk 348 has a quite flat spectrum up to the millimetre wavelengths 
(S$_\nu \sim \nu^{-0.09}$).  
The fact that the broadband spectrum is smooth and flat between the radio 
and millimetre
bands suggests that the emission up to the millimetre band is synchrotron
emission. The wavelength at which the non-thermal emission has a cut-off
occurs at wavelengths shorter than 1.35 mm, perhaps between the sub-millimetre
and far-infrared bands. The SED of Mrk 348 is very similar to that of
other radio-loud objects, in which the
radio emission is thought to be the result, either directly or indirectly,
of the presence of a relativistic jet. We will discuss this further in the 
next section.\\

\subsection{Radio Polarization}
\noindent The polarisation seen in
the southern component in Mrk348 is unusual as most Seyferts  show
little radio polarisation. This lack of polarised flux is readily
explained by the de-polarising effect of the ionised gas in the narrow
line region(NLR), within which the radio emission is usually embedded. The
fact that the southern component is 5\% polarised, whereas the
stronger central component is unpolarised at 6 cm ($<0.5\%)$, can be most
easily interpreted by assuming the southern component to be in front of
the NLR gas, whereas the central component is embedded within it. This
would be consistent with models where the southern component is associated
with a jet pointing towards the observer. Although it could be argued that
the lack of polarisation in the central and northern components is
intrinsic, models in which the southern component is on the far side of
the NLR can almost certainly be ruled out. In fact a similar situation is
seen in NGC1068 \cite{wilulv87}, where only the NW lobe shows
significant polarisation, and the central and southern components are
depolarised by NLR gas.
This effect is analogous, albeit on a much smaller scale,  to the
Laing-Garrington effect \cite{laing88,garrington88} seen in radio
galaxies.

\section{Discussion}
\label{mrk348radiodiscussion}

\noindent One of the main reasons for studying Mrk 348 is that the
observational evidence appears to point to different and mutually
inconsistent scenarios. Our aim is to try and come up with a single
model which is consistent with all the observational facts. This
involves close examination of the evidence concerning the orientation
of the AGN with respect to the line of sight and whether or not
Doppler boosting is playing a dominant role in determining its
observed properties. The amount of beaming has a strong bearing on the
degree of radio loudness of the object; is Mrk 348 radio quiet or is it one
of those (relatively rare) objects intermediate between radio-loud
and radio-quiet objects?

\subsection{The orientation of Mrk 348}
\begin{figure}
\centerline{\psfig{figure=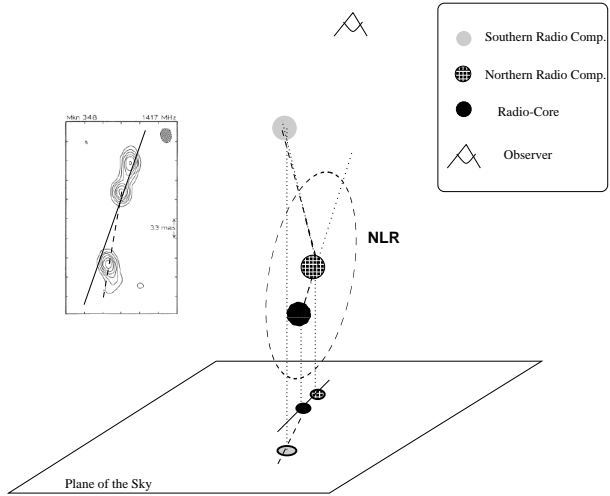,width=8cm}}
\caption{A possible geometrical configuration for Mrk 348, see description in
the text}
\label{projeccaook}
\end{figure}

\noindent The NOT images of Mrk 348 clearly show that the AGN is
hosted by a giant nearly face-on spiral galaxy. The near circular
symmetry of the galaxy is preserved down to the ring of HII regions on
a scale of $\sim$1~kpc. The continuum radio properties of Mrk 348 such
as the flat radio spectrum continuing well into the infrared without
any sign of a break, the core-dominated radio structure and the rapid
radio variability \cite{neff83}, are all consistent with the source
axis making a small angle to the line of sight to the observer. On the
other hand the optical data, for example the Seyfert 2 spectrum and
scattered polarized broad lines, all support a different scenario in
which the axis of the AGN makes a large angle to the line of sight to
the observer. The detection of strong H$_2$O megamaser emission in
Mrk 348 by Falcke et al. (2000) might be taken as evidence to
reinforce this conclusion since megamasers are almost exclusively
found in Seyfert 2 galaxies which are believed to have edge-on molecular
tori. However, the maser emission consisting of a single reshifted
broad-line is not like that seen from the prototype edge-on disk
object NGC4258 (Miyoshi et al. 1995) which has both red- and
blue-shifted narrow lines. Both Peck et al. (2001) and Xanthopoulos \&
Richards (2001), using VLBA and MERLIN observations respectively, have
reached the conclusion that the maser emission probably arises from
molecular material shocked by the jet and not from a torus. Since the maser emission is
only seen towards the northern VLBI jet, and it is redshifted with
respect to the systemic velocity of the galaxy, the shock model
suggests that the northern jet is pushing the molecular material away
from us and thus implies that the northern jet is the receding one.
Beyond this, it seems that the detection of H$_2$O maser emission in Mrk 348
places no unambiguous constraint on the orientation of the AGN
axis with the line of sight to the observer.\\ 
Tran (1995) argues on
the basis of his spectropolarimetry that the torus is being viewed at
a low inclination. By the same argument as above this would
support a scenario where the AGN axis makes a small angle to the line
of sight to the observer.  The evidence for ionization cones is also
relevant to the orientation debate, but the data do not seem to be
conclusive: a) Simpson et al. (1996) claim to see ionization cones in
Mrk 348, from ground-based images, with an half-angle $\sim$ 45~$\deg$
b) the evidence for such cones in high resolution (HST) images
\cite{fal98} is, however, marginal. \\
\noindent The VLBI results on Mrk 348 of U99 are potentially
important. U99 detect a one-sided core-jet and measure
sub-relativistic motions in the jet with $\beta_{app}$=0.08. Again,
the evidence is pointing in two different directions; the one-sided
nuclear jet and the more symmetric larger-scale structure seen in the
MERLIN maps, could naturally be accounted for by beaming of a
relativistic nuclear jet but the sub-luminal motion fits more
naturally into a picture with the jets making a larger angle to the
line of sight \footnotemark\footnotetext{ There are examples of
quasars with some knots showing superluminal motions, but which also
have knots that are apparently stationary, e.g. 3C 395
\cite{lara94}. U99 believe, nevertheless, that the measured speeds
represent the true jet speeds as the core and northern structures have
not changed between the (two) epochs.}. (We point out that a
naive combination of motion of 0.08c (24 000~$km s^{-1}$) in the plane
of the sky and the radial motion of 130~$kms^{-1}$, deduced from the maser 
emission, indicates a resultant motion
almost completely in the plane of the sky.) U99 suggest that the absence
of a southern counterpart to the north-pointing jet could be accounted
for by free-free absorption by ionised gas with average density $n_e >
2\times10^5$cm$^{-3}$, resulting in a column density of $n_H\sim$
10$^{23}$cm$^{-2}$. This value is completely consistent with the value
calculated from X-ray observations \cite{war89}.  We note, however,
that \scite{gallimore99} do not detect HI absorption in front of any
of the radio components and they estimate an upper limit of $n_H\sim$
5$\times10^{19}$cm$^{-2}$.  That is, the X-ray and radio observations
give very different $n_H$ suggesting that the detected X-ray and
radio-core radiation is propagating through regions with different
opacities -- perhaps because the bases of the radio emitting jets are
located further out along the AGN axis than the X-ray emitting
regions. If this is the case, the similarity of the required $n_H$ to
free-free absorb the southern component and the calculated $n_H$ (from
X-ray observations) is a coincidence.\\
\noindent The MERLIN polarization observations compound the confusion;
attributing the polarization asymmetry to a Laing-Garrington-like effect
\cite{laing88,garrington88} in which the near-side of the source
suffers less Faraday depolarization than the far-side, argues that the
southernmost component is the closest to the observer. This,
incidentally, is consistent with the shock model of the maser emission
put forward by Peck et al. (2001) and Xanthopoulos \& Richards
(2001).  The U99 VLBI results, however, show only a north-pointing
jet, suggesting that the north-side is the approaching side of the
source. This is true whether the absence of an observable southern
counter-jet is a result of beaming or free-free absorption by a
disk.  Can this apparent conflict be resolved?  It could be that in
Mrk 348, instead of the obvious hypothesis that we are seeing a
lobe-core-lobe structure, we are actually seeing a single one-sided
jet that is moving towards the observer. In Figure \ref{projeccaook}
we present a cartoon of a possible geometrical configuration. Here,
the core and ``northern'' components are further inside the host
galaxy, and the ``southern'' component represents the end of the jet,
further outside the host galaxy. In our model, both ``northern'' and
``southern'' components are moving towards the observer. The direction
of the jet makes a small angle with the line of sight, but then bends
by a small angle. Projection effects magnify the small bending, making
the ``southern'' component appear at the south of the core.  In this
situation the true counter-jet might not be detected, in agreement
with the U99 findings, if it is dimmed by Doppler de-boosting
effects. \\
\noindent We summarize our conclusions about orientation as follows:

\begin{itemize}

\item The optical data  seems to point  to Mrk 348 fitting 
the standard model for a
Seyfert 2 galaxy in which there is an obscuring structure hiding the
optical AGN from view. This requires axis of the system to make a
large ($\geq45\deg$) angle to the line of sight.

\item The radio core-dominated structure with small-scale one-sided
jet would most naturally fit a model in which the radio jet axis makes
a small angle ($\leq10\deg$) to the line of sight and in which Doppler
boosting plays an important role. However, the lack of detectable
superluminal motion has led U99 to suggest that the angle to the line
of sight is not small and propose free-free absorption by a disk to
account for the one-sided VLBI structure and sub-luminal motion. But
the polarization asymmetry would then still remain without an obvious
explanation.

\item The H$_2$O megamaser emission adds little to the orientation
debate apart suggesting that, if the motion megamaser gas is
associated with that of the jet, the northerly-pointing jet is receding 
from us.

\item Our provisional conclusion is to take the optical data at face
value and assume in the subsequent discussion that the angle to the line
of sight is large, and that Doppler beaming is not a dominant factor.

\end{itemize}

\subsection{Mrk 348, a radio-quiet or a radio-loud object?}

Radio-loud objects are generally defined as objects in which the ratio
between the flux density at 5 GHz and the flux density at B band is
S$_5$/S$_{\mbox{\tiny B}}\ge10$ \cite{kel89}. We now consider
the radio flux density. If Doppler beaming plays at most a minor role, 
then the observed radio flux density should be close to the intrinsic 
value.  We now look at the optical flux density.  We note that the 
S$_5$/S$_{\mbox{\tiny B}}$
definition takes no explicit account of the dilution of the AGN light
by background starlight \cite{bro93}, which can be important in the
case of luminous host galaxies. This means that S$_5$/S$_{\mbox{\tiny
B}}$ can be an underestimate of the object's nuclear radio activity.
\noindent Nevertheless, we can estimate the optical nuclear flux
density of Mrk 348 at B band from the strength of the 4000\AA\ break
in a similar way to that of March\~a et al. (1996).  The ratio gives
S$_5$/S$_{\mbox{\tiny B}}$=933, a value that puts Mrk 348 in the range
of radio-loud sources. On the other hand, S$_{\mbox{\tiny B}}$
represents a lower limit to the true flux density since it is only the
light which is scattered into our line of sight which is detected. In
this situation the detected central continuum S$_c$ -- the fraction
that is scattered towards the observer direction -- represents a
fraction of the true central flux S$_T$.  We note that it would be
necessary that $S_c$ $< 0.01 S_T$, i.e. a fraction of scattered light
smaller than 1\%, to raise S$_{\mbox{\tiny B}}$ by a factor of 100,
and put S$_5$/S$_{\mbox{\tiny B}}\sim 10$ (i.e.  at the
radio-loud/radio-quiet boundary). Nevertheless it has been found that
the typical fraction of scattered light is $\sim$ 2\%\ (R. Fosbury in
``Portugal-ESO-VLT'' meeting, 2000). Thus, the ratio
S$_5$/S$_{\mbox{\tiny B}}$ suggests that Mrk 348 is close to, but
slightly above, the radio-loud/radio-quiet boundary. In this context the
combination of radio loudness and a spiral host galaxy is unusual. 
The above discussion assumes that Doppler beaming plays at most a minor
role. But if Doppler beaming is important then  Mrk 348 is 
intrinsically a radio-quiet object. In this context the combination
of its radio-loudness and host galaxy type would fit nicely the 
expectation for a radio-intermediate quasar.

\section{Summary}
\label{section-conclusions}

\noindent Mrk348 has been classified as a Seyfert galaxy. One
peculiarity of this source is that it is a relatively strong radio
object hosted by a spiral galaxy, not by an elliptical galaxy. 
 In this paper we have presented NOT U, B, R, and I
images of Mrk 348. From profile fitting calculations we conclude that
the host galaxy of Mrk 348 is between Sa+ and Sb+ morphological type.
On large scales, there is evidence that Mrk348 and
its eastern companion are interacting, but based on the view of the
symmetry of the inner region of Mrk348 we do not think that this
interaction has triggered the Mrk 348 nuclear activity.
The peculiarity of Mrk 348 comes from the SED and the polarisation seen in 
the southern component: both are unusual amongst radio-quiet objects.
We have discussed the  orientation of Mrk 348 with respect to the line of 
sight. The evidence points in different directions, some observations are
 consistent
with the axis of the system to be at large  ($\geq45\deg$) angle to the 
line of sight, whereas other observations are consistent if the 
radio jet axis makes a small ($\leq10\deg$) to the line of sight.
The orientation of Mrk 348 with the line of sight is a
difficult matter to resolve and it might be that in Mrk 348 the
geometry is much more complicated than a simple edge-on or face-on
model. If the radio
emission in Mrk348 is beamed then Mrk 348
is intrinsically a radio-quiet object. Detection of superluminal
motions would be the ultimate test of such a model. Note that recently
superluminal motions were detected in III Zw 2, a Seyfert galaxy that
had not previously shown superluminal motions in the past \cite{Brunthaler00}.
But if the angle to the line
of sight is large, and if Doppler beaming is not a dominant factor,
Mrk 348 is close to, but slightly above, the radio-loud/radio-quiet boundary.
In this case, the peculiarity of Mrk 348 comes from the combination of
a radio-loud object harboring a spiral host galaxy.

\section{Acknowledgments}
We thank our referee Dr Heino Falcke for constructive comments. The ISOPHOT data were reduced during a visit to the ISOPHOT Data
Centre.  S\'onia Ant\'on acknowledges all the support received during this
visit, in particular that from Martin Haas. S\'onia Ant\'on
acknowledges the European Commission, TMR Programme, Research Network
Contract ERBFMRXCT96-0034 ``CERES'', and PRAXIS XXI Programme through
the grant BD/5532/95. The NOT is operated on the island of La Palma
jointly by Denmark, Finland, Iceland, Norway, and Sweden, in the
Spanish Observatorio del Roque de los Muchachos of the Instituto de
Astrofisica de Canarias.

\end{document}